\documentclass[prb,showpacs,twocolumn,aps,superscriptaddress,a4paper]{revtex4-1}
\usepackage{dcolumn,amssymb,amsmath,amsfonts,graphicx,latexsym,color,braket}

\definecolor{myblue}{rgb}{0,0,0.75}

\begin{document}

\title{A theory of nonequilibrium steady states in quantum chaotic systems}

\author{Pei Wang}
\email{wangpei@zjnu.cn}
\affiliation{Department of Physics, Zhejiang Normal University, Jinhua 321004, China}

\date{\today}

\begin{abstract}
Nonequilibrium steady state (NESS) is a quasistationary state, in which exist currents that
continuously produce entropy, but the local observables are stationary everywhere.
We propose a theory of NESS under the framework of quantum chaos.
In an isolated quantum system whose density matrix follows a unitary evolution,
there exist initial states for which
the thermodynamic limit and the long-time limit are noncommutative.
The density matrix $\hat \rho$ of these states displays a universal structure.
Suppose that $\ket{\alpha}$ and
$\ket{\beta}$ are different eigenstates of the Hamiltonian with energies $E_\alpha$ and $E_\beta$, respectively.
$\bra{\alpha}\hat \rho \ket{\beta}$ behaves as a random number which has zero mean.
In thermodynamic limit, the variance of $\bra{\alpha}\hat \rho \ket{\beta}$
is a smooth function of $\left| E_\alpha-E_\beta\right|$,
scaling as $1/\left| E_\alpha-E_\beta\right|^2$ in the limit $\left| E_\alpha-E_\beta\right|\to 0$.
If and only if this scaling law is obeyed, the initial state evolves into NESS in the long time limit.
We present numerical evidence of our hypothesis in a few chaotic models.
Furthermore, we find that our hypothesis indicates the eigenstate thermalization hypothesis (ETH)
for current operators in a bipartite system.
\end{abstract}

\maketitle

\section{Introduction}


A unified theory of nonequilibrium steady state (NESS) is still unaccessible
up to now in spite of intense study in statistical mechanics~\cite{zwanzig}.
According to second law of thermodynamics, a macroscopic isolated system
will eventually relax to thermal equilibrium~\cite{landau}.
But if the system is infinitely large, it may take infinitely long time
to remove the imbalances in the initial state. The system then can never
reach thermal equilibrium. Instead, it will relax to a nonequilibrium steady state~\cite{evans}, in which
exist currents that continuously produce entropy,
but the local observables are stationary everywhere.
For example, a system composing of two infinite
reservoirs at different temperatures will relax to NESS, in which the
heat persistently flows from the hotter reservoir to the colder one.

In this paper we discuss the theory of NESS in quantum systems.
NESS in classical systems is also an active area of research~\cite{zwanzig}.
Nevertheless, the microscopic mechanical law is quantum.
Various approaches have been developed for studying NESS~\cite{datta}.
Kubo formula~\cite{kubo} is valid if the deviation from thermal equilibrium is infinitesimal.
The nonequilibrium Green's functions technique~\cite{rammer} was employed in the irreversible
processes starting from an equilibrium state. Landauer-B\"{u}ttiker
formula~\cite{buttiker85,buttiker86} was specifically designed
for a scattering region coupled to multiple thermal reservoirs.
Hershfield~\cite{hershfield93} derived an expression for the density matrix of NESS.
These approaches all depend on an explicitly defined initial state.
But statistical mechanics reminds us that
the initial memory is lost in the thermalization process.
And a thermalized state depends only upon
very few parameters like the total energy and particle number.
The idea of lost memory is at the heart of maximized entropy principle,
which lays a foundation of the unified description of equilibrium states.
Similarly, one expects that some redundant information in the initial state
should be lost in the evolution to NESS.
And a unified description of NESS emerges once if
the surviving information can be distinguished from the lost information.

The lost information in the thermalization process has been well addressed,
thanks to the development of quantum chaos theory.
According to this theory, generic (i.e. chaotic) systems with complicated interactions between
particles must be distinguished from integrable systems.
The former can thermalize, but the latter cannot~\cite{eisert}.
The eigenstate thermalization hypothesis (ETH) was proposed~\cite{deutsch,srednicki94,srednicki99}.
It explains why an isolated system
loses its memory in spite of the fact that the wave function follows a unitary evolution.
ETH states that the matrix elements of physical observables in
the eigenbasis of Hamiltonian can be expressed as~\cite{srednicki99,alessio15}
\begin{equation}\label{eq:eth}
 O_{\alpha \beta} = O(\bar E) \delta_{\alpha,\beta} + e^{- S (\bar E )/2} f_O(\bar E,\epsilon) R_{\alpha \beta},
\end{equation}
where $\bar E=(E_\alpha +E_\beta)/2$ and $\epsilon=E_\alpha-E_\beta$ denote the
average of and the difference between two eigenenergies, respectively.
The diagonal element $O(\bar E)$ is a smooth function of energy. While the off-diagonal elements
are exponentially small with $S(\bar E )$ denoting the thermodynamic entropy.
$S$ is related to the density of many-body states by $D = e^{S}$.
The off-diagonal elements are the product of a smooth function
$f_O(\bar E,\epsilon)$ and a random number $R_{\alpha \beta}$ with zero mean
and unit variance. Starting from a typical initial state, the long time limit of observables
depends only upon the main diagonal of the initial density matrix~\cite{rigol08}
which is sometimes called the diagonal ensemble.
The diagonal ensemble is not necessarily an equilibrium ensemble.
But according to Eq.~(\ref{eq:eth}), one cannot distinguish the values of observables
with respect to different eigenstates whose energies are the same. Therefore,
the diagonal ensemble and the equilibrium ensemble
predict same results for the observables. In this sense a chaotic quantum system thermalizes.
Strictly speaking, no information of the initial density matrix is destroyed under a unitary evolution.
However, it is impossible to extract this information from local observables, whose
expectation values depend only upon few parameters like the system's total energy and particle number.
In this sense, the initial memory is lost.
The loss of memory is hidden behind the fact
that the off-diagonal elements of the density matrix average out in the thermalization process
and that the local observables depend only upon the eigenenergy.

If a chaotic system evolves into NESS, one expects that the initial memory
should be lost in a similar way. But ETH cannot explain the existence of NESS by itself.
To address the nature of NESS, we propose the nonequilibrium steady state hypothesis (NESSH).
This hypothesis provides a unified description of NESS and clarifies
which information in the initial state is lost in the evolution to NESS.

\section{Definition of nonequilibrium steady states}

\begin{figure}[tbp]
\includegraphics[width=0.9\linewidth]{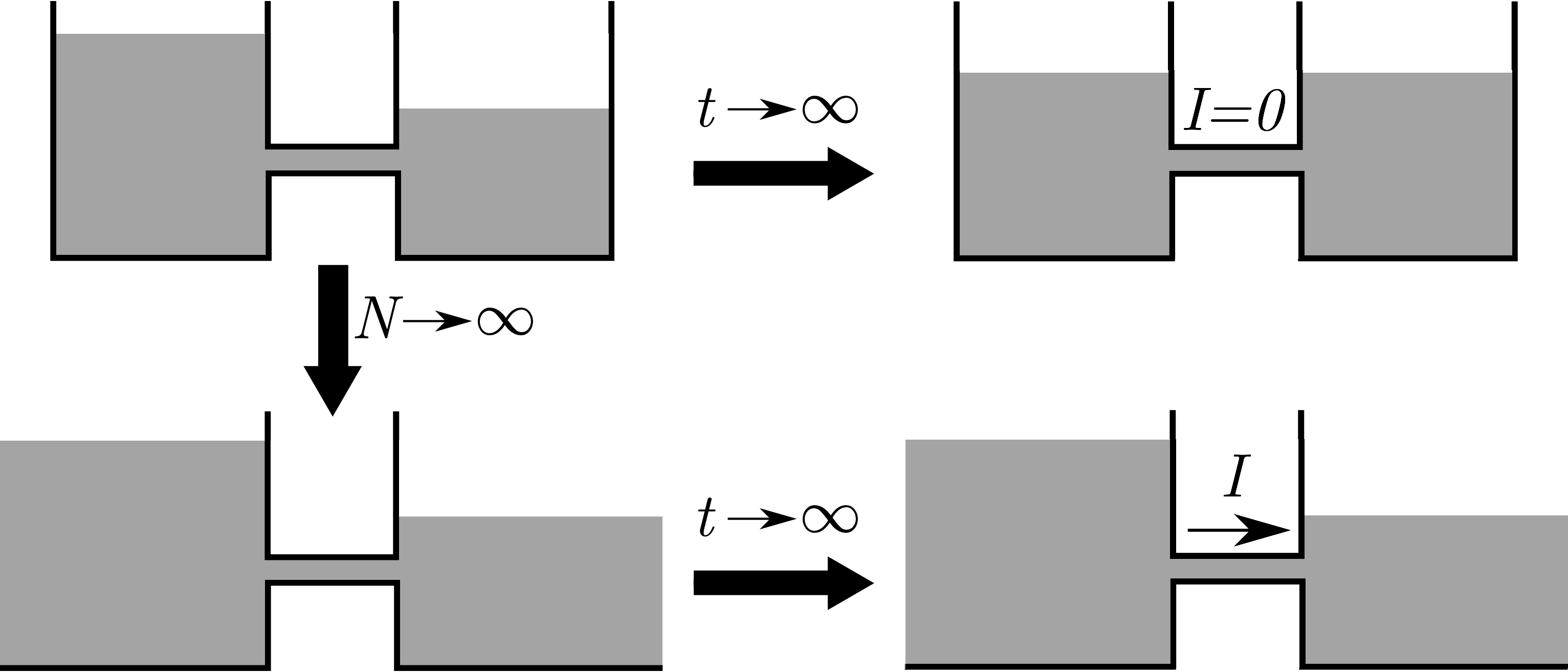}
\caption{Schematic diagram of a bipartite system hosting NESS. The
long time limit $t\to\infty$ and the thermodynamic limit $N\to \infty$ are noncommutative.}\label{fig:schematic}
\end{figure}
Let us first discuss the definition of NESS in isolated systems.
Some authors are used to talking about NESS in open systems. Nevertheless, an open system
can always be treated as part of a larger isolated system.

We notice next facts:

(1) The exclusive characteristic of NESS is the existence of ``nonequilibrium'' currents.
These currents are distinguished from the circular currents
that may exist in some equilibrium states, e.g., the persistent current in a mesoscopic ring~\cite{buttiker83}.
Nonequilibrium currents result from the tendency to remove the particle (energy) distribution imbalance.
The system returns to thermal equilibrium by removing these imbalances.
Therefore, nonequilibrium currents produce entropy.
For example, the heat flow from the hotter part of a system to the colder part is a nonequilibrium current.
Integrable systems after a quench will relax to non-thermal states~\cite{rigol06}
in which there is no current or entropy production.
We distinguish these non-thermal states from NESS.
The system is in NESS if and only if there exist stationary nonequilibrium currents,
which is denoted by $\hat I$. To obtain $\hat I$ of a specific model, one divides
the isolated system into parts. The current flowing into or out of one part
is expressed as the time derivative of the corresponding extensive quantity
(e.g. energy or particle number) of this part, which can be further expressed
as the commutator between the extensive quantity and the Hamiltonian according to the Heisenberg equation.

(2) In a {\bf finite} isolated system, eigenstates do not carry nonequilibrium currents.
Because if there exist nonequilibrium currents, the particle (energy)
distribution must change with time due to the conservation law. This contradicts the fact
that physical observables keep invariant in an eigenstate. Especially, let us consider a bipartite system
with the number of particles in the left and right part being $\hat N_L$ and $\hat N_R$, respectively.
The nonequilibrium current from $L$ to $R$ is $\hat I = d \hat N_{R} /dt = i [\hat H, \hat N_R]$.
It is straightforward to prove $\bra{\alpha} \hat I \ket{\alpha}=0 $ for an eigenstate $\ket{\alpha}$.
Note that eigenstates in finite systems can carry circular currents,
since they do not change the particle (energy) distribution.

(3) In a finite system, if an observable relaxes to its stationary value
in the long time limit, it must be determined by the diagonal ensemble~\cite{rigol08}:
\begin{equation}\label{eq:argdiagonalensemble}
 \begin{split}
  \lim_{t\to \infty} O(t) = & \lim_{T\to \infty} \frac{1}{T} \int_0^T dt O(t)
  = \textbf{Tr} \left[ \hat \rho_{d} \hat O\right],
 \end{split}
\end{equation}
where $\hat \rho_{d} = \sum_\alpha \bra{\alpha}\hat \rho\ket{\alpha} \ket{\alpha}\bra{\alpha}$
is the diagonal ensemble. $\hat \rho$ is the initial density matrix.
Nonequilibrium currents must be zero in the diagonal ensemble due to fact~(2).
Therefore, diagonal ensemble is distinguished from NESS.

Due to facts~(1-3), NESS should be a quasistationary state. In a finite system,
this state survives for a while, but finally relaxes to the diagonal ensemble.
But if the system's size goes to infinity, i.e. the thermodynamic limit,
the lifetime of this quasistationary state goes to infinity. It then becomes a real steady state.
NESS exists if and only if the thermodynamic limit $N\to \infty$ and the long time limit $t\to \infty$
are noncommutative:
\begin{equation}\label{eq:noncommutative}
 \lim_{N\to \infty} \lim_{t\to\infty} I(t) \neq  \lim_{t\to\infty} \lim_{N\to \infty} I(t).
\end{equation}
According to fact~(3), the left hand side equals zero. Therefore,
the right hand side is nonzero, i.e., a stationary nonequilibrium current survives.

We define NESS as follows:
{\bf If the thermodynamic limit and the long time limit
are noncommutative for a specific initial state of an isolated system, taking $N\to \infty$ before
taking $t \to \infty$ results in a NESS.} It is worth emphasizing that in NESS the two limits
are noncommutative for arbitrary observables, but not only for the current.

An example is helpful for understanding the noncommutativity of the two limits.
Let us consider two reservoirs at different water levels which
are connected by a pipe (see Fig.~\ref{fig:schematic}).
Water flows from left to right to remove the level imbalance.
The flow stops after the left and right levels reach the same.
But if the reservoirs are infinitely large, the flow never stops, since
the level imbalance cannot be removed within finite period.
Fig.~\ref{fig:schematic} also indicates how the thermodynamic limit
is taken. Each infinite reservoir is described by a few intensive variables including
the temperature and the chemical potential.
We must keep all these intensive variables invariant
when taking the limit $N\to \infty$.

When a system evolves into NESS, two different situations must be distinguished.
If the system is integrable, there is no universal description for NESS,
which has to be studied model by model by solving the Schr\"{o}dinger equation under certain initial conditions.
On the other hand, the nonintegrable systems share some common features,
e.g., their local observables can always be expressed as Eq.~(\ref{eq:eth})
independent of whether the system is a gas, liquid or solid or which particles the system is made of.
ETH proposes an assumption about the observable operator to explain
why a nonintegrable system thermalizes.
Similarly, we expect an assumption to explain why a nonintegrable system evolves into NESS.
Different from ETH, such an assumption should be about the structure of the density matrix, as shown in next.
Our motivation is to obtain a general description of NESS,
just as the Gibbs ensemble is a general description of equilibrium states.

\section{Nonequilibrium steady state hypothesis}
\label{sec:nessh}

Let us start from the Schr\"{o}dinger equation and see why Eq.~(\ref{eq:noncommutative}) is possible.
The time-dependent current is expressed as
\begin{equation}\label{eq:current}
 I(t) = \sum_{\alpha\neq \beta}e^{-i(E_\alpha-E_\beta)t} \rho_{\alpha\beta}
I_{\beta \alpha},
\end{equation}
where $\rho_{\alpha\beta}= \bra{\alpha}\hat \rho \ket{\beta}$ and
$I_{\beta \alpha}=\bra{\beta}  \hat I \ket{\alpha}$ are
the initial density matrix and the current matrix, respectively.
$\alpha$ and $\beta$ are the eigenstates of the Hamiltonian.
$E_\alpha$ and $E_\beta$ are their eigenenergies, respectively.
Notice that terms with $\alpha =\beta$ are excluded from the sum, because the diagonal
elements of $I$ are zero.
If $\displaystyle\lim_{t\to\infty} I(t)$ exists, it must be
equal to the averaged current over $[0,T]$ as $T\to \infty$,
i.e. $\bar I= \displaystyle\sum_{\alpha\neq \beta} \displaystyle\frac{e^{-i(E_\alpha-E_\beta)T}-1}
{-i(E_\alpha-E_\beta)T} \rho_{\alpha\beta} I_{\beta \alpha} $. At a specific $T$,
the pairs of eigenstates with energy difference $\left| E_\alpha-E_\beta \right| \gg 2\pi/T$ do not
contribute to $\bar I$. We say that the phase coherence between these pairs is lost.
If the system's size is finite,
the level spacing has a minimum, therefore, there always exists sufficiently large $T$
when all the eigenstate pairs satisfy $\left| E_\alpha-E_\beta \right| \gg 2\pi/T$.
The phase coherence is totally lost, and the stationary current must be zero.
But if the system's size is infinite, the level spacing goes to zero.
For arbitrarily large $T$, there exist eigenstate pairs that satisfy $\left| E_\alpha-E_\beta \right| \ll 2\pi/T$.
The phase coherence between these pairs survives. The superposition of these neighbor eigenstates
may carry a finite nonequilibrium current as $\bra{\beta} \hat I \ket{\alpha}\neq 0$.
In above we do not consider the degeneracy, which is broken in a chaotic system.
Strictly speaking, the level spacing in chaotic systems follows the well-known
Wigner-Dyson distribution~\cite{wigner,dyson} which vanishes at zero if the system's size is finite.
As the system's size goes to infinity, the peak of Wigner-Dyson distribution
shifts towards zero and the averaged level spacing decays to zero exponentially.

NESS is essentially a partially-coherent state, which exists in thermodynamic limit.
In a finite system with the averaged level spacing $\Delta$,
the phase coherence between the eigenstate pairs is totally lost at the time scale $\hbar/ \Delta$.
While the current relaxes to a stationary value at a much shorter time
which is denoted as $\hbar/\Gamma$. $\Gamma$ is determined by the interaction
strength or the bandwidth of the system.
During the time $\hbar/\Gamma \ll t \ll \hbar/ \Delta$, the current is quasistationary.
This quasistationary current approaches the steady current in NESS as the system's size
goes to infinity.
One can study the properties of NESS by doing
a proper scaling in finite systems~\cite{wang11}.

Chaotic systems have some universal properties.
According to ETH~(\ref{eq:eth}), the current matrix has
indeed a general expression in an arbitrary chaotic system, which is
\begin{equation}\label{eq:currenteth}
 I_{\alpha \neq \beta} =  e^{- S (\bar E )/2} f_I(\bar E,\epsilon) R^I_{\alpha \beta}.
\end{equation}
Note that $\bar E=(E_\alpha +E_\beta)/2$ and $\epsilon=E_\alpha-E_\beta$.
We then guess that the off-diagonal elements of the density matrix have a similar expression:
\begin{equation}\label{eq:ansatz1}
 \rho_{\alpha \neq \beta} = e^{- S (\bar E )/2}
 f_\rho(\bar E,\epsilon) R^\rho_{\alpha \beta}. \tag{i}
\end{equation}
Ansatz~(\ref{eq:ansatz1}) is distinguished from ETH.
Because the density matrix is not a few-body operator,
and then must be distinguished from physical observables.
The diagonal element $\rho_{\alpha\alpha}$ is not necessarily a smooth function of $E_\alpha$.
Notice that ansatz~(\ref{eq:ansatz1}) stands for a generic state, but not for a fine-tuned one
such as eigenstates. We note that, to the best of our knowledge,
ansatz~(\ref{eq:ansatz1}) has not been clearly written down before,
but the idea behind it is not new. In random matrix theory (RMT), it was proved that
the eigenstates of random matrices in any basis are random unit vectors~\cite{alessio15}.
Due to the similarity between random matrices and quantum chaotic systems,
a generic state in the eigenbasis of a chaotic system should also be a random vector.
Correspondingly, the off-diagonal elements of a generic density matrix are random numbers.
Ansatz~(\ref{eq:ansatz1}) goes further by proposing an
envelop function $f_\rho(\bar E,\epsilon)$.

The randomness of $\rho_{\alpha\beta}$ and $I_{\alpha\beta}$
can be understood as follows. Choose two real numbers $\bar{E}$ and $\epsilon$.
If the system's size is large enough, there should be many eigenstate pairs $(\alpha,\beta)$
whose averaged energy $(E_\alpha+E_\beta)/2$ and energy difference $E_\alpha-E_\beta$
fall within the thin shells centered at $\bar{E}$ and $\epsilon$, respectively.
The value of $ \rho_{\alpha\beta}$ or $I_{\alpha\beta}$ fluctuates
with $(\alpha,\beta)$ like a random number with some proposed distribution.
Notice that in thermodynamic limit,
the number of pairs within a shell goes to infinity.
After taking the thermodynamic limit we
should set the shell width to infinitesimal for obtaining the distribution of
$\rho_{\alpha\beta}$ or $I_{\alpha\beta}$. Because in general the distribution
depends on $\bar{E}$ and $\epsilon$.

It is worth emphasizing that $R^I_{\alpha \beta}$ and $R^\rho_{\alpha \beta}$ are
two random numbers with zero mean and unit variance.
Their correlation is defined as the average of their product
over a thin shell surrounding specific values of $\bar{E}$ and $\epsilon$:
\begin{equation}\label{eq:corr}
\mathcal{C}_{\rho I} = \overline{ R^\rho_{\alpha\beta}R^I_{\beta\alpha} }.
\end{equation}
$\mathcal{C}_{\rho I}$ is a function of $\bar{E}$ and $\epsilon$.
But it is reasonable to suppose that $\mathcal{C}_{\rho I}$ changes slowly with $\bar{E}$ or $\epsilon$,
so that in the calculation of current
we can treat $\mathcal{C}_{\rho I}$
as the constant $\mathcal{C}_{\rho I}({\bar{E}=E,\epsilon=0})$
where $E$ denotes the total energy of the system (see the discussion below).

Since the entropy $S(\bar E)$ is real, the hermitianity of $\rho$ and $I$ requires
\begin{eqnarray}
\begin{array}{cc}
f_\rho (\bar E, \epsilon) =  f_\rho^* (\bar E, -\epsilon), & R^\rho_{\alpha\beta} = \left(R^{\rho}_{\beta \alpha}\right)^*, \\
f_I (\bar E, \epsilon) =  f_I^* (\bar E, -\epsilon), & R^I_{\alpha\beta} = \left(R^{I }_{\beta \alpha}\right)^*.
\end{array}
\end{eqnarray}
Let us consider a popular situation - a real Hamiltonian together with
a purely imaginary current operator. This corresponds to
a system with time-reversal symmetry. For example, let us consider a fermionic lattice model
with the Hamiltonian $\hat H= -\sum_{m,j}g_{m,j} \hat c^\dag_m \hat c_j +
U_{m,j} \hat n_m \hat n_j$ where $\hat n_j = \hat c^\dag_j \hat c_j$. The current operator from site $m$ to $j$
is $\hat I = i g_{m,j} \left( \hat c^\dag_{j} \hat c_m - \text{H.c.} \right) $.
Since $\rho_{\alpha\beta}$ is real but $I_{\alpha\beta}$ is purely imaginary,
we suppose $R^I$ and $R^\rho$ to be real symmetric matrices. And we suppose
$f_\rho (\bar E, \epsilon) =  f_\rho (\bar E, -\epsilon)$ to be real
but $f_I (\bar E, \epsilon) =  -f_I (\bar E, -\epsilon)$ to be purely imaginary functions.

Now let us discuss the condition of $f_\rho$ under which
the nonequilibrium current survives in the steady limit.
Substituting Eq.~(\ref{eq:currenteth}-\ref{eq:corr}) and ansatz~(\ref{eq:ansatz1})
into Eq.~(\ref{eq:current}), we obtain
\begin{equation}\label{eq:currexp}
\begin{split}
 I(t)= \mathcal{C}_{\rho I} \int_{-\infty}^\infty d\bar{E} D(\bar{E}) \int_{-\infty}^\infty
 d\epsilon e^{-i\epsilon t} f_\rho (\bar E, \epsilon)
 f_I(\bar E, - \epsilon),
 \end{split}
\end{equation}
where $D(\bar{E})=e^{S(\bar{E})}$ denotes the density of many-body states. To get Eq.~(\ref{eq:currexp})
we used $\sum_{\alpha} \to \int dE_\alpha D\left(E_\alpha\right)$
and $D\left( \bar{E}\pm \epsilon/2\right) \approx D\left(\bar{E}\right)$.
The latter approximation is due to the fast decay of $f_\rho (\bar E, \epsilon)
 f_I(\bar E, - \epsilon)$ as $\left|\epsilon\right|$ increases.
Therefore, most contribution to $\int d\epsilon$
comes from a small interval centered at $\epsilon=0$ in which
$D\left( \bar{E}\pm \epsilon/2\right) \approx D\left(\bar{E}\right)$.
It was already observed that the off-diagonal elements of observables
decay quickly as $\left|\epsilon\right|$ increases~\cite{khatami}.
Eq.~(\ref{eq:currexp}) is surprisingly
simple. The detail of the model and the initial state is hidden behind
the random matrices $R^I$ and $R^\rho$. Their contribution to
$I(t)$ is simplified into a constant $\mathcal{C}_{\rho I}$.
$\displaystyle \lim_{t\to \infty} I(t) \neq 0$
is equivalent to $\displaystyle\lim_{t\to\infty} \int_{-\infty}^\infty d\epsilon e^{-i\epsilon t} f_\rho 
 f_I \neq 0 $, i.e., the Fourier transformation of $f_\rho f_I$
is nonzero in the limit $t\to\infty$.
According to Riemann-Lebesgue lemma,
the Fourier transformation of an integrable function decays to zero in the limit $t\to \infty$.
Therefore, $f_\rho f_I$ cannot be an integrable function.
Instead, $f_\rho f_I $ must be the product of $1/\epsilon$ and some integrable function.
In fact, one can derive
$\displaystyle \lim_{t\to \infty} I(t) \neq 0$ and $\displaystyle \lim_{t\to \infty} dI(t)/dt = 0$
from this assumption (see Appendix~\ref{sec:app} for more detail).
The existence of a steady nonequilibrium current
requires that either $f_\rho$ or $f_I $ diverges as $1/\epsilon$ in the limit $\epsilon\to 0$.
In fact, it is $f_\rho$ which diverges as $1/\epsilon$
if the initial state evolves into a NESS, but $f_I $ always converges to a finite value.
Otherwise, if $f_I $ diverges as $1/\epsilon$,
we have $\displaystyle\lim_{t\to\infty} I(t)\neq 0$ for arbitrary initial state,
which contradicts the fact that only a part of initial states
relax to NESS but the others thermalize ($\displaystyle\lim_{t\to\infty}I(t)=0$).

According to the above argument, we propose our second ansatz.
Initial states can be classified into typical and atypical states.
Typical initial states thermalize in the long time limit, i.e. $\displaystyle \lim_{t\to\infty} I(t)=0$.
The envelop function $f_\rho$ of typical initial density matrices is
an integrable function of $\epsilon$.
On the other hand, atypical initial states evolve into NESS, i.e. $\displaystyle \lim_{t\to\infty} I(t)\neq 0$.
The corresponding envelop function scales as $1/\epsilon$ in the limit $\epsilon\to 0$.
One can express the atypical envelop function as
\begin{equation}\label{eq:ansatz2}
 f_\rho (\bar E, \epsilon) = \frac{\rho(\bar{E},\epsilon)}{\left| \epsilon \right|}, \tag{ii}
\end{equation}
where $\rho(\bar{E},\epsilon)$ converges to a finite value in the limit $\epsilon\to 0$.
The denominator is $\left|\epsilon\right|$
because the hermitianity requires $f_\rho (\bar E, \epsilon)$ to be even.
Ansatz~(\ref{eq:ansatz1}) and~(\ref{eq:ansatz2}) combine into our
nonequilibrium steady state hypothesis: {\bf The off-diagonal elements of an initial density matrix that evolves into NESS
can be expressed as}
\begin{equation}\label{eq:nessh}
 \rho_{\alpha \neq \beta} = 
 \frac{e^{- S (\bar E )/2}\rho(\bar{E},\epsilon)}{\left| \epsilon \right|} R^\rho_{\alpha \beta}. \tag{NESSH}
\end{equation}
NESSH means that the off-diagonal elements of atypical density matrices
are random numbers. And their variance scales as $1/\epsilon^2$ in the diagonal limit.

Eq.~(\ref{eq:nessh}) is an expression for the density matrix of a system of finite size.
Note that the density matrix by itself does not have a well-defined thermodynamic limit.
The thermodynamic limit is taken in the calculation of local observables, e.g. the current.
This is done as we transform the sum over $\alpha,\beta$
into the integral over $\epsilon$ in Eq.~(\ref{eq:currexp}). If the system's size is finite,
the domain of $\epsilon$ must have an infrared cutoff - the level spacing $\Delta$.
We take the thermodynamic limit by neglecting this cutoff and
setting the domain of $\epsilon$ to $\left(-\infty,\infty\right)$.
Note that the density of states $D=e^{S}$ increases exponentially with
the system's size. But $\rho$ and $f_\rho$ scale as $e^{-S}$, that is decrease exponentially
with the system's size. While $\mathcal{C}_{\rho I}$
and $f_I$ converge as the system's size goes to infinity.
Therefore, the thermodynamic limit of $D(\bar{E})f_\rho(\bar{E},\epsilon)$ and
then the integrand in Eq.~(\ref{eq:currexp}) are well defined.
One can also separate the factor $e^{-S}$ from $\rho (\bar{E}, \epsilon)$,
which would only cause a difference of notation.

According to Eq.~(\ref{eq:currexp}) and ansatz~(\ref{eq:ansatz2}),
if $\rho (\bar E, \epsilon)$ is real but $f_I(\bar E, \epsilon)$ is purely imaginary,
the expression of stationary current can be further simplified into
\begin{equation}\label{eq:currsimexp}
\begin{split}
\lim_{t\to\infty} I(t)= i \pi \mathcal{C}_{\rho I} \int_{-\infty}^\infty d\bar{E} D(\bar{E}) 
\rho (\bar E, 0)f_I(\bar E, 0^+),
 \end{split}
\end{equation}
where $f_I(\bar E, 0^+)=\displaystyle\lim_{\epsilon\to 0^+} f_I(\bar E, \epsilon)$.
Notice that the odd function $f_I(\bar E, \epsilon)$ is discontinuous
at $\epsilon =0$. This will be verified in the following numerical experiments.
To get Eq.~(\ref{eq:currsimexp}) we used the Dirichlet integral $\int dx \sin(x)/x =\pi$.

In previous study, it has been observed that
the diagonal elements of the density matrix $\rho_{\alpha\alpha}$
as a function of $E_\alpha$ is Gaussian-like. And its
variance is sufficiently small, i.e. behaving like in traditional
statistical mechanics ensembles~\cite{alessio15}. Similarly, we suppose that
the off-diagonal element $\rho_{\alpha\beta}$ as a function of $\bar{E}$ is also Gaussian-like with a
small variance $\delta \bar{E}$. Therefore, $\rho(\bar{E},0)$ has a
sharp peak of width $\delta \bar{E}$ centered at $E=\textbf{Tr}(\hat \rho \hat H)$
which is just the total energy of the system. It is reasonable to
suppose that $\mathcal{C}_{\rho I}$, $D$ and $f_I$ all change slowly
in the range $(E-\delta \bar{E}, E+\delta \bar{E})$ so that they can
be treated as constants. Finally, the expression of current becomes
\begin{equation}\label{eq:currsimplestexp}
\begin{split}
\lim_{t\to\infty} I(t)= i \pi \mathcal{C}_{\rho I} D(E) 
\rho (E, 0)f_I(E, 0^+).
 \end{split}
\end{equation}
This expression is valid in an arbitrary chaotic system.

Starting from NESSH, we can prove for an arbitrary
observable $\hat O$ that $\displaystyle \lim_{t\to\infty} O(t)$ exists and is distinguished
from the value of $\hat O$ in the diagonal
ensembles, i.e. $\sum_\alpha \rho_{\alpha\alpha} O_{\alpha\alpha}$.
This meets our definition of NESS. $O(t)$ can be divided into $\sum_\alpha \rho_{\alpha\alpha} O_{\alpha\alpha}$
and $O_{NE}(t)$. The latter comes from
the off-diagonal elements of initial density matrices.
$O_{NE}(t)$ can be obtained in the same way as $I(t)$.
According to NESSH, we have
\begin{equation}\label{eq:one}
O_{NE}(t)= \mathcal{C}_{\rho O} \int_{-\infty}^\infty d\bar{E} D(\bar{E}) \int_{-\infty}^\infty
 d\epsilon e^{-i\epsilon t} \frac{\rho(\bar{E},\epsilon)}{\left| \epsilon \right|}
 f_O(\bar E, - \epsilon)
\end{equation}
with $\mathcal{C}_{\rho O}=\overline{ R^\rho_{\alpha\beta}R^O_{\beta\alpha} }$.
Let us study the derivative $dO_{NE}/dt $, which
is the Fourier transformation of $\rho f_O$. But
$\rho$ and $f_O$
are both integrable functions. According to Riemann-Lebesgue lemma,
in the limit $t\to\infty$, $dO_{NE}/dt $ vanishes
and then $O_{NE}$ must approach a stationary value.
And this stationary value is nonzero since the integrand in Eq.~(\ref{eq:one})
is not integrable with $\left|\epsilon\right|$ appearing in the denominator.
The limit of $O_{NE}(t)$ is just the commutator between
the thermodynamic limit and the long time limit:
\begin{equation}\label{eq:otnoncomm}
\lim_{t\to\infty} O_{NE}(t) =
\lim_{t\to\infty} \lim_{N\to \infty} O(t)-\lim_{N\to \infty} \lim_{t\to\infty} O(t).
\end{equation}
To see why Eq.~(\ref{eq:otnoncomm}) is true, we
start from $O(t)=\sum_{\alpha}O_{\alpha\alpha}\rho_{\alpha\alpha}
+\displaystyle\sum_{\alpha\neq \beta} e^{-i(E_\alpha-E_\beta)t} \rho_{\alpha\beta}
O_{\beta \alpha}$. Here the first (second) term is called the diagonal (off-diagonal)
term. Different from $I(t)$, $O(t)$ may have a nonzero diagonal term.
For the time-independent diagonal term, the two limits $N\to\infty$ and
$t\to \infty$ are commutative.
On the other hand, the off-diagonal term vanishes if we take $t\to \infty$ first
(see Eq.~(\ref{eq:argdiagonalensemble}) and the surrounding argument).
But the off-diagonal term survives if $N\to\infty$ is taken first,
and its value is just what we define as $O_{NE}(t)$. Eq.~(\ref{eq:otnoncomm})
is thus obtained.

NESSH clarifies how the initial memory is lost in the evolution to NESS.
The density matrix follows a unitary evolution:
\begin{equation}\label{eq:unitarydensity}
\rho_{\alpha\beta}(t)= \delta_{\alpha,\beta}\rho_{\alpha\alpha}
+ e^{-i\epsilon t} \frac{e^{- S (\bar E )/2}
\rho(\bar{E},\epsilon)}{\left| \epsilon \right|} R^\rho_{\alpha \beta}.
\end{equation}
The information of the initial density matrix
cannot be destroyed under a unitary evolution. But most of these information
cannot be extracted from the local observables such as the current.
In the thermalization process, no information contained in the second term of Eq.~(\ref{eq:unitarydensity})
can be extracted. Or we say these information are all lost.
NESS keeps more information than thermalized states.
The stationary current depends on the value of $\rho(\bar{E},\epsilon)$ in the limit $\epsilon\to 0$.
The off-diagonal elements with $\left|\epsilon\right|>0$
average out in the evolution to NESS.
NESS only keeps memory of the off-diagonal elements with infinitesimal energy difference.
Furthermore, $\rho_{\alpha\alpha}$ is insensitive to the change of $\alpha$
once if $E_\alpha$ is fixed (ETH). This explains why thermalization happens.
Similarly, the detail of the initial state is contained in the matrix
$R^\rho_{\alpha \beta}$. But $R^\rho_{\alpha \beta}$ contributes to
the value of observables through its correlation with $R^O_{\alpha \beta}$.
Physical observables are then insensitive to the detail of $R^\rho_{\alpha \beta}$.
This is the reason why NESS looks ``universal''.

\section{Numerical experiments in random matrices}

We test NESSH (ansatz~(\ref{eq:ansatz1}) and~(\ref{eq:ansatz2})) in a few chaotic models.

Let us first consider a bipartite structure as shown in Fig.~\ref{fig:schematic}.
The system composes of two weakly-coupled reservoirs
(the meaning of ``weakly-coupled'' will be discussed below).
The Hamiltonian of each reservoir is a random matrix or to be specific, a
Gaussian orthogonal ensemble (GOE)~\cite{kravtsov}.
In detail, GOE is a real symmetric matrix.
Its diagonal (off-diagonal) entries are independent random numbers,
and each follows the Gaussian distribution with zero mean and variance $\sigma^2$ ($\sigma^2/2$).
We have two reasons for choosing random matrices. First, random matrices
are believed to have the same properties as quantum chaotic systems.
Second, NESSH should in principle be tested in thermodynamic limit.
While random matrices of small dimensions already display
thermodynamic properties. To see these properties in ``real'' models,
the dimensions of the Hamiltonian have to be very large and the numerical
calculation is therefore more difficult. Anyway, we also test NESSH
in a ``real'' model. The results will be discussed in next section.

Suppose that there are $n$ eigenstates in each reservoir. The eigenenergies are denoted as
$\varepsilon_1, \varepsilon_2, \cdots, \varepsilon_n$.
According to random matrix theory, the probability density of eigenenergies is~\cite{kravtsov}
\begin{equation}\label{eq:rmtleveldis}
 P(\varepsilon_1, \varepsilon_2, \cdots, \varepsilon_n)= \frac{e^{-\frac{\varepsilon_1^2+\cdots+\varepsilon_n^2}{2\sigma^2}}
 \left| \prod_{m>j}\left(\varepsilon_m -\varepsilon_j\right) \right|}
 {\sigma^{\frac{n(n+1)}{2}}\left(2\pi\right)^{\frac{n}{2}} \prod_{j=1}^n \displaystyle\frac{\Gamma\left(1+j/2\right)}{\Gamma(3/2)}}.
\end{equation}

We use $\gamma_L$ and $\gamma_R$ to denote the eigenstates of the left and right reservoir, respectively.
They are not the eigenstates of the whole system, since the two reservoirs are coupled.
The coupling Hamiltonian is expressed as a matrix $V$
in the basis $\ket{\gamma_L \gamma_R}$. The matrix elements
$V_{\gamma_L \gamma_R,\gamma'_L \gamma'_R}$ are independent
random numbers. Each follows the Gaussian distribution with zero mean and variance $\sigma_c^2$.
For the coupling to be weak, we require $\sigma_c \ll \sigma$.
Furthermore, $\sigma_c$ must change with the system's size while $\sigma_c n$ keeps a constant.
This scaling behavior can be understood by considering next example.
Two chains of length $l$ are coupled at the end sites.
The single-particle Hamiltonian of each chain can be diagonalized by a Fourier transformation,
which results in $l$ single-particle levels in each chain.
After the transformation, the rescaled coupling between left and right single-particle levels
must have an extra factor $1/l$. Or $l$ times the coupling strength is a constant.
This condition guarantees that the coupling energy does not increase with the system's size.
The energy flow between reservoirs is then bounded as the reservoir's size goes to infinity.
This is necessary for the initial imbalance not being removed in finite period.
NESS can only exist under the weak coupling condition.

The total Hamiltonian can be expressed as
\begin{equation}\label{eq:rmthamiltonian}
\begin{split}
 \hat H=& \sum_{\gamma_L,\gamma_R} \left(\varepsilon_{\gamma_L}+ \varepsilon_{\gamma_R}\right)
 \ket{\gamma_L \gamma_R} \bra{\gamma_L\gamma_R}\\ & + V_{\gamma_L \gamma_R,\gamma'_L \gamma'_R}
 \ket{\gamma_L \gamma_R} \bra{\gamma'_L\gamma'_R}.
 \end{split}
\end{equation}
The eigenstate of the whole system is denoted as $\ket{\alpha}$ which satisfies $\hat H\ket{\alpha}
=E_\alpha \ket{\alpha}$. We employ $\ket{\gamma_L\gamma_R}$ as the initial state.
This corresponds to that the two reservoirs are initially decoupled and the coupling
is then switched on for the heat to flow.
The initial imbalance manifests as the difference between $\varepsilon_{\gamma_L}$
and $\varepsilon_{\gamma_R}$. Without loss of generality, we set $\epsilon_{\gamma_L} > \epsilon_{\gamma_R}$,
i.e. the left reservoir is hotter than the right one.
If $\left|\varepsilon_{\gamma_L}-\varepsilon_{\gamma_R}\right|$
increases with the system's size and goes to infinity in thermodynamic limit,
the initial imbalance will survive in the long time limit.
We denote the inner product between the initial state and the eigenstate
as $K^\alpha_{\gamma_L\gamma_R}
=\braket{\gamma_L\gamma_R | \alpha}$.
NESSH should then be equivalently expressed as
\begin{equation}\label{eq:rmtnessh}
 K^\alpha_{\gamma_L\gamma_R}K^\beta_{\gamma_L\gamma_R} = 
 \frac{e^{- S (\bar E )/2}\rho(\bar{E},\epsilon)}{\left| \epsilon \right|} R_{\alpha \beta}.
\end{equation}
Note that Eq.~(\ref{eq:rmtnessh}) is for a system of finite size,
in which case $ K^\alpha_{\gamma_L\gamma_R}K^\beta_{\gamma_L\gamma_R}$
as a function of $\epsilon= E_\alpha -E_\beta$ has no singularity since $\epsilon$
has an infrared cutoff - the level spacing $\Delta$.

\subsection{NESSH indicates ETH for the current operator}

Let us study the current operator.
Here the nonequilibrium current is the energy current between two reservoirs.
Due to the conservation of total energy, we define the current from left to right as
$\hat I= - d \hat H_L(t)/dt$. $\hat H_L=\sum \epsilon_{\gamma_L} \ket{\gamma_L}\bra{\gamma_L}$
denotes the Hamiltonian of the left reservoir. The matrix elements of $\hat I$ in the eigenbasis are
$ I_{\alpha\neq \beta} = - i \epsilon  \displaystyle\sum_{\gamma_L\gamma_R} \varepsilon_{\gamma_L}
 K^\alpha_{\gamma_L\gamma_R}K^\beta_{\gamma_L\gamma_R}$.
Substituting Eq.~(\ref{eq:rmtnessh}) in, we immediately obtain
\begin{equation}\label{eq:currrmt}
I_{\alpha\neq \beta} = - i \textbf{sgn} (\epsilon) e^{- \frac{S (\bar E )}{2}}
\sum_{\gamma_L\gamma_R} \varepsilon_{\gamma_L} \rho(\bar{E},\epsilon) 
R_{\alpha \beta}.
\end{equation}
Since $R_{\alpha \beta}$ for different $\left(\gamma_L\gamma_R\right)$
are independent random numbers and
$\rho$ is an integrable function, the sum of $\varepsilon_{\gamma_L} \rho
R_{\alpha \beta}$
should also be an integrable function times a random number with zero mean and unit variance.
We define $f_IR^I_{\alpha\beta} = - i \textbf{sgn} (\epsilon)
\sum \varepsilon_{\gamma_L} \rho R_{\alpha \beta}$. Eq.~(\ref{eq:currrmt})
is then just the eigenstate thermalization hypothesis~(\ref{eq:currenteth}).
In this way, we showed that NESSH indicates ETH for the current operator. Notice that we have not
used the proposition that each reservoir is described by a random matrix.
Our derivation stands in arbitrary bipartite systems.
Note that NESSH is a statement about the density matrix of the quantum state,
while ETH is a statement about the observable operators. They are two different
statements for chaotic systems. The above analysis demonstrates the relation between them.

Furthermore, $\textbf{sgn}(\epsilon)$ appears in the expression of $I_{\alpha\beta}$,
indicating that the odd function $f_I(\bar E, \epsilon)$ is discontinuous
at $\epsilon =0$, as what we expected.
The correlation $\mathcal{C}_{\rho I}$ can be extracted from Eq.~(\ref{eq:currrmt}), which is
\begin{equation}
 \mathcal{C}_{\rho I} = - i \textbf{sgn} (\epsilon)
\frac{ \varepsilon_{\gamma_L} \rho(\bar{E},\epsilon) }{f_I(\bar{E},\epsilon)}.
\end{equation}
Here we used $\overline{R_{\alpha\beta}R_{\beta\alpha}}=1$.

\subsection{NESSH in $2$-by-$2$ random matrices}

Let us consider the few body limit - only two levels in the left
reservoir and a single level in the right one. The total Hamiltonian is a $2$-by-$2$ matrix:
\begin{eqnarray}
\hat H = \left( \begin{array}{cc}
           H_{11} & H_{12} \\
           H_{21} & H_{22}
          \end{array} \right).
\end{eqnarray}
Here $H_{12}=H_{21}$ is a random number denoting the coupling
between reservoirs. The two eigenstates of $\hat H$ are denoted as $\alpha=\left(\alpha_1,\alpha_2\right)^T$
and $\beta=\left(\beta_1,\beta_2\right)^T$. The corresponding eigenenergies
are $E_\alpha$ and $E_\beta$, respectively. With some boring but straightforward calculation, we
can express the eigenvectors in terms of $H_{12}$, $E_\alpha$ and $E_\beta$. We then obtain
\begin{equation}
 \alpha_1\beta_1= - \alpha_2\beta_2 = \frac{H_{12}}{\left|E_\alpha-E_\beta\right|}.
\end{equation}
This is just the NESSH~(\ref{eq:rmtnessh}). It means that the off-diagonal elements of
initial density matrix is a random number with the variance scaling as $1/\left|E_\alpha-E_\beta\right|^2$.

Note that the distribution of $H_{12}$ is not precisely Gaussian in case of
fixed $E_\alpha$ and $E_\beta$. The joint probability $ P\left(E_\alpha,E_\beta,H_{12}\right) $
is indeed
\begin{equation}
\begin{split}
&  P\bigg(H_{11}\left(E_\alpha,E_\beta,H_{12}\right),H_{22}\left(E_\alpha,E_\beta,H_{12}\right)\bigg) \\ & \times
 \frac{e^{-H_{12}^2/\left(2\sigma_c^2\right) }}{\sqrt{1-\frac{4H_{12}^2}{\left(E_\alpha-E_\beta\right)^2}}},
 \end{split}
\end{equation}
where $P(H_{11},H_{22})$ is given by Eq.~(\ref{eq:rmtleveldis}) in case of $n=2$.
One can prove that, $H_{12}$ approximately follows a Gaussian
distribution with the constant variance $\sigma_c^2$ once if $\left(E_\alpha-E_\beta\right)^2 \gg 4H^2_{12}$.
This condition is equivalent to the weak coupling condition.

\subsection{NESSH in thermodynamic limit}

After discussing the case of $2$-by-$2$ matrix,
we turn to the thermodynamic limit which NESSH is proposed for.
We numerically diagonalize the Hamiltonian~(\ref{eq:rmthamiltonian})
of dimensions up to tens of thousands. This corresponds to about $100$ energy levels in
each reservoir. We verify the hypothesis~(\ref{eq:rmtnessh}) in two steps. First, we
show that $ K^\alpha_{\gamma_L\gamma_R}K^\beta_{\gamma_L\gamma_R}$ is a random number and plot
its distribution.
Second, we show that the variance of $ K^\alpha_{\gamma_L\gamma_R}K^\beta_{\gamma_L\gamma_R}$
becomes a smooth function of $\epsilon=E_\alpha-E_\beta$ in thermodynamic limit.
And it scales as $1/\epsilon^2$ for small $\left|\epsilon\right|$.

\begin{figure}[tbp]
\includegraphics[width=0.9\linewidth]{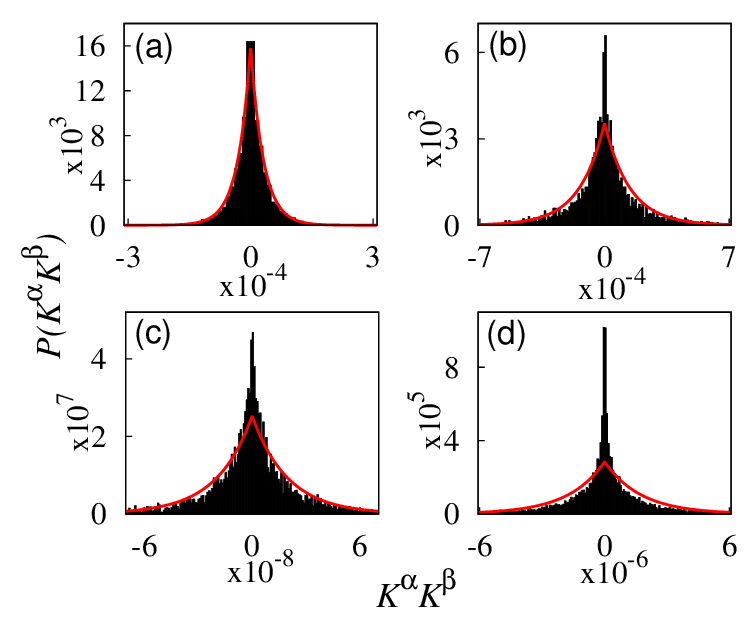}
\caption{(Color online) The distribution of $\rho_{\alpha\beta}$ at $\left|E_\alpha-E_\beta\right|=3$
in a bipartite random-matrix model ((a) and (b)) and in a lattice model
((c) and (d)). $(E_\alpha+E_\beta)/2=0$ is at the middle of the band.
The red line shows the Laplace distribution with the same variance.
(a) $n=100$ and $\sigma_c=0.01\sigma$. $\gamma_L$ ($\gamma_R$)
is the $10$th highest (lowest) level. $K_u$ evaluates $4.4$. (b) $\sigma_c=0.1\sigma$.
The other parameters are the same as (a). $K_u$ evaluates $3.8$. (c) $n_f=6$, $n_r=8$
and $U_1=U_2=0.8$. $K_u$ evaluates $7.9$. (d) $U_1=U_2=0.5$. The other
parameters are the same as (c). $K_u$ evaluates $11.2$.}\label{fig:distribution}
\end{figure}
Recall that the randomness of $ K^\alpha_{\gamma_L\gamma_R}K^\beta_{\gamma_L\gamma_R}$
should be estimated in a set of $\left(\alpha,\beta\right)$ with
$\left(E_\alpha+E_\beta\right)/2$ and $E_\alpha-E_\beta$ falling within
thin shells centered at $\bar{E}$ and $\epsilon$, respectively.
The shell width should be infinitesimal in thermodynamic limit.
In the numerical simulation, we set the shell width to
approximately two orders of magnitude smaller than the bandwidth of reservoirs.
It is small enough for the influence of shell width on the distribution being neglected.
While the shell still contains a few thousands samples, many enough for the
distribution being correctly displayed. The average of samples is found to be zero, fitting our prediction.
We calculate the standard deviation of samples, which is denoted as $\sigma_s$.
We then divide the interval $[-3\sigma_s,3\sigma_s]$ into $200$ bins and count
the number of samples falling in each bin. The histogram is plotted in Fig.~\ref{fig:distribution}.
Note that the $y$-axis is rescaled for the integral of
$P\left(K^\alpha_{\gamma_L\gamma_R}K^\beta_{\gamma_L\gamma_R}\right)$ being normalized to unity.
Fig.~\ref{fig:distribution}(a) and~(b) show the probability density
$P\left(K^\alpha_{\gamma_L\gamma_R}K^\beta_{\gamma_L\gamma_R}\right)$
at different coupling strength. It looks regular and is symmetric to zero, as we expected.

To further study the property of this distribution,
we calculate the excess kurtosis of samples.
The excess kurtosis of a random number $X$ with zero mean is defined as
\begin{equation}
K_u = \frac{\overline{X^4}}{\left(\overline{X^2}\right)^2} - 3.
\end{equation}
The result of $K_u$ changes with model parameters. It is close to but larger than $K_u=3$.
Note that $K_u$ of the Laplace distribution~\cite{laplace} is exactly $3$.
In Fig.~\ref{fig:distribution}(a) and (b) we compare the distribution of samples with the Laplace
distribution of the same variance (the red lines).
Their shapes look similar to each other but the difference is also clear.
The distribution of samples has a sharper peak and lower shoulders.

\begin{figure}[tbp]
\includegraphics[width=0.9\linewidth]{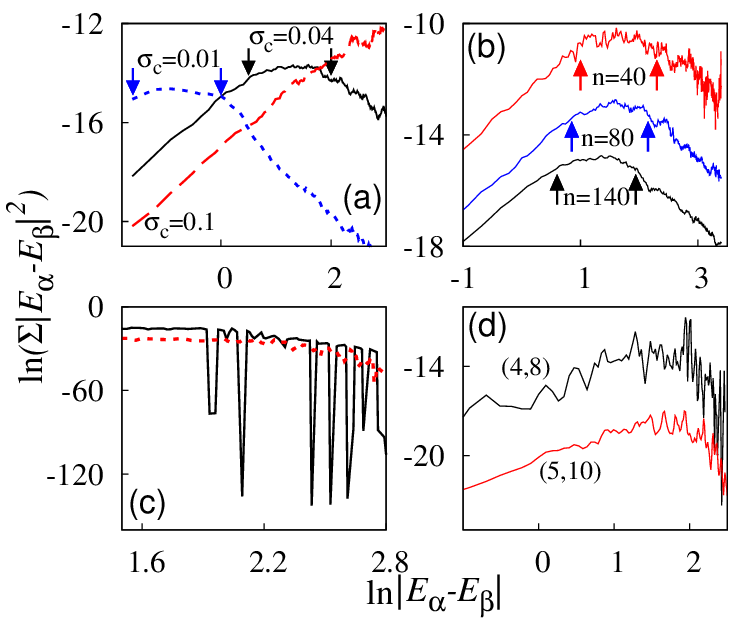}
\caption{(Color online) The variance of $\rho_{\alpha\beta}$ times $(E_\alpha-E_\beta)^2$
as a function of $\left|E_\alpha-E_\beta\right|$ in the logarithmic scale.
(a) and (b) are for the random-matrix model, and (c) and (d) are for the lattice model.
The arrows mark the start and end points of the plateau. (a) The variance at different
coupling is compared, as the system's size is fixed to $n=100$.
(b) The variance at different system's size is compared, as the coupling
is fixed to $n\sigma_c=4$.
(c) The chaotic region ($U_1=U_2=0.5$) is compared with the integrable region
($U_1=U_2=0$). (d) The variance at different system's size is compared,
as the particle density is fixed to $n_f/n_r=1/2$. The black
line is for $n_f=4$ while the red one for $n_f=5$.}\label{fig:variance}
\end{figure}
In thermodynamic limit, NESSH predicts that the variance of
$K^\alpha_{\gamma_L\gamma_R}K^\beta_{\gamma_L\gamma_R}$
scales as $1/\left(E_\alpha-E_\beta\right)^2$ in the limit
$\left|E_\alpha-E_\beta\right|\to 0$.
Arbitrary atypical initial states must obey this scaling law.
Let us see what happens in a finite system.
In a finite system, $\left|E_\alpha-E_\beta\right|$
has a lower bound - the level spacing $\Delta$.
The variance scales as $1/\epsilon^2$ in a range $(\epsilon_-,\epsilon_+)$.
Correspondingly, the nonequilibrium quasistationary state survives in the
period $\hbar/\epsilon_+ < t < \hbar/\epsilon_-$.
At the beginning of Sec.~\ref{sec:nessh}, we have argued
that $\hbar/\Gamma \ll t \ll \hbar/\Delta$. We then have
$\epsilon_- > \Delta$. This means that the variance might deviate from $1/\epsilon^2$
in a finite system as $\epsilon\to \Delta$.
However, $\epsilon_-$ must approach zero in thermodynamic limit.
Because the quasistationary state becomes a real steady state in thermodynamic limit.
On the other hand, the variance for {\it typical} initial states
does not display the $1/\epsilon^2$ scaling behavior. Instead,
it is convergent in the diagonal limit.

We denote the variance of
$K^\alpha_{\gamma_L\gamma_R}K^\beta_{\gamma_L\gamma_R}$ as $\Sigma$.
To address the scaling behavior of $\Sigma$,
we plot $\Sigma\epsilon^2$ as a function of $\epsilon$ in the logarithmic scale
(see Fig.~\ref{fig:variance}).
This function should be a constant if $\Sigma\sim 1/\epsilon^2$.
Therefore, a plateau in $\Sigma \epsilon^2$ (or in $\ln\left(\Sigma \epsilon^2\right)$)
signals the NESSH scaling law.

Fig.~\ref{fig:variance}(a) plots $\ln\left(\Sigma \epsilon^2\right)$ vs.
$\ln\epsilon$ for different coupling strength.
The curve for $\sigma_c=0.01\sigma$ (weak coupling) is significantly
distinguished from that for $\sigma_c=0.1\sigma$ (strong coupling).
The former displays a clear plateau at small $\epsilon$, verifying
the NESSH scaling law.
At $\sigma_c=0.04\sigma$, the plateau is also clear but its position
moves to the middle of the domain.
But the curve for $\sigma_c=0.1\sigma$ has a slope of $2$
in almost the whole domain of $\epsilon$, i.e. $\Sigma$ keeps a constant.
In the strong coupling regime (the coupling increases with the system's size),
the two reservoirs are in fact a unity.
The energy flow then has no upper bound in thermodynamic limit.
The initial imbalance can always be removed in finite period.
Therefore, all the initial states are typical states and will thermalize in the long time limit.
$\Sigma$ being a constant in the strong coupling regime
is consistent with our theory.
In fact, atypical states or NESS can be only found in the weak coupling regime.

In Fig.~\ref{fig:variance}(b) we compare $\Sigma \epsilon^2$
at different system's size. As the system's size increases, the fluctuation of $\Sigma \epsilon^2$
is suppressed. We then expect $\Sigma$ to be a smooth function of $\epsilon$
in thermodynamic limit.
And the plateau shifts towards smaller $\epsilon$ as the system's size increases,
indicating $\epsilon_- \to 0$ in thermodynamic limit.
The numerical results fit with the prediction of NESSH.
Furthermore, $\Sigma$ becomes independent of $\epsilon$ to the left of the plateau (smaller $\epsilon$),
but decays as $1/\epsilon^4$ to the right of the plateau (larger $\epsilon$).
Note that $\Sigma \propto f_\rho^2$.
According to the expression of current~(\ref{eq:currexp}),
a constant $\Sigma$ at small $\epsilon$ implies that the current decays to zero
at large $t$. While $\Sigma \sim 1/\epsilon^4$ (or $f_\rho \sim 1/\epsilon^2$) at large $\epsilon$
implies that the current changes linearly at small $t$.
Because the second derivative of $\int d\epsilon e^{-i\epsilon t} f_I(\epsilon) /\epsilon^2$
with respect to $t$ is $\int d\epsilon e^{-i\epsilon t} f_I(\epsilon)$,
which quickly decays to zero as $t$ increases.
The second derivative being zero indicates that the first derivative, i.e. $dI/dt$
is a constant, or the current changes linearly.
From the shape of $\Sigma(\epsilon)$ we deduce that, after the coupling between reservoirs is switched on,
$I$ first increases linearly to its quasistationary value, stays at this value
for a while, and then decays to zero.
This behavior meets our expectation.

\section{Numerical experiments in a lattice model}

\begin{figure}[tbp]
\includegraphics[width=0.9\linewidth]{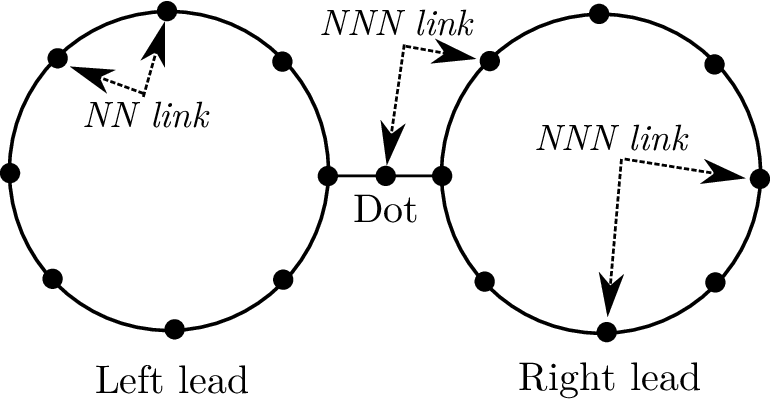}
\caption{Schematic diagram of the spinless fermionic model
with $n_r=8$. Some of the nearest-neighbor (NN)
links or the next-nearest-neighbor (NNN) links are marked
by a pair of arrows.}\label{fig:schfermi}
\end{figure}
Above random-matrix model
does not originate from the microscopic description of matter.
To verify NESSH in a realistic model, we consider the spinless fermions
located on a lattice of shape ``$\infty$'' (see Fig.~\ref{fig:schfermi} for the schematic
diagram). This model is a simplified version
of a quantum dot coupled to two metallic leads.
The left and right circles simulate
the left and right leads, respectively. Each contains $n_r$ sites.
We employ a circle shape to avoid the scattering at the boundary.
The two circles are connected via a center site (the quantum dot).
The lattice then has $2n_r+1$ sites in total.
The fermions are only permitted to hop between the nearest-neighbor sites.
If two fermions simultaneously occupy two nearest-neighbor sites, the interaction energy is $U_1$.
If they occupy two next-nearest-neighbor sites, the interaction energy is $U_2$.
The total Hamiltonian is
\begin{equation}\label{eq:Hfermion}
\hat H= -\sum_{\langle i,j\rangle} \left(\hat c^\dag_i \hat c_j + \text{H.c.}\right) +
U_1 \sum_{\langle i,j\rangle} \hat n_i \hat n_j + U_2 \sum_{\langle\langle i,j\rangle\rangle} \hat n_i \hat n_j,
\end{equation}
where $\hat c^\dag_i$ and $\hat c_j$ denote the fermionic creation and
annihilation operator, respectively, $\hat n_i=\hat c^\dag_i \hat c_i$
denotes the on-site occupation, and $\langle i,j\rangle$ and
$\langle\langle i,j\rangle\rangle$ denote a pair of nearest-neighbor
and next-nearest-neighbor sites, respectively.

The total number of fermions is denoted as $n_f$ which satisfies $n_f<n_r$.
The initial imbalance is realized by putting all the fermions on the left circle.
To keep it simple, the initial position of fermions is random.
We then estimate the distribution of the off-diagonal elements $\rho_{\alpha\beta}$
in the initial density matrix.
The distribution at different $\left(U_1,U_2\right)$ is plotted in Fig.~\ref{fig:distribution}(c)
and (d).
Again, the distribution is symmetric to zero and
has a sharp peak at zero. The distribution decays quickly as $\rho_{\alpha\beta}$
deviates from zero. The shape of this distribution is similar to that of the Laplace distribution,
but has a much sharper peak. The value of $K_u$ is larger than $3$.

We estimate the variance of $\rho_{\alpha\beta}$ as a function of $\epsilon$.
The results are plotted in Fig.~\ref{fig:variance}(c) and (d).
The Hamiltonian~(\ref{eq:Hfermion}) is nonintegrable (chaotic) if $U_1$ and $U_2$ are nonzero,
but is integrable at $U_1=U_2=0$. This provides us a chance for checking the
difference between chaotic systems and integrable systems.
For integrable systems, the variance $\Sigma$
depends strongly on the choice of the energy shell and does not
have a well-defined thermodynamic limit.
With fixed shell width,
the curve $\Sigma$ vs. $\epsilon$ displays a strong fluctuation (see Fig.~\ref{fig:variance}(c), the black curve).
This fluctuation is not suppressed as the system's size increases.
But for chaotic systems, the fluctuation of $\Sigma$ is much weaker
(see Fig.~\ref{fig:variance}(c), the red curve).
And Fig.~\ref{fig:variance}(d) shows that
the fluctuation becomes even weaker as the system's size increases.
In thermodynamic limit, $\Sigma$ should be a smooth function of $\epsilon$.
Therefore, NESSH only stands in a chaotic system.

A plateau in the function $\Sigma\epsilon^2$ can be observed (see Fig.~\ref{fig:variance}(d)),
indicating the emergence of the NESSH scaling law.
Unfortunately, up to the largest system's size that we can handle,
we do not find the trend of the plateau moving towards smaller $\epsilon$.
Different models should be considered in future study.

\section{Conclusions}

Let us summarize the main ansatz and results of our theory.
NESS is a quasistationary state in finite systems,
in which it will eventually relax to thermal equilibrium.
But in thermodynamic limit, NESS is a real steady state,
because the thermodynamic limit and the long time limit are noncommutative.
The initial states can be classified into typical and atypical states.
Typical states thermalize in the long time limit, while
atypical states will evolve into NESS.
NESSH proposes the universal structure of atypical states in chaotic systems,
as they are expressed as density matrices in the eigenbasis of the Hamiltonian.
As shown in Eq.~(\ref{eq:nessh}) of Sec.~\ref{sec:nessh} (or Eq.~(\ref{eq:ansatz1})
and~(\ref{eq:ansatz2})),
the off-diagonal elements $\rho_{\alpha\beta}$ of atypical density matrices
behave as random numbers. Their variance is a smooth function of $\left|E_\alpha-E_\beta\right|$,
scaling as $1/\left|E_\alpha-E_\beta\right|^2$ in the limit $\left|E_\alpha-E_\beta\right|\to 0$.
This scaling law is the exclusive characteristic of NESS.

Based on this ansatz, the stationary current in NESS can be simply expressed as
the variance of $\rho_{\alpha\beta}\left|E_\alpha-E_\beta\right|^2$ in the limit
$\left|E_\alpha-E_\beta\right|\to 0$ (see Eq.~(\ref{eq:currsimplestexp})).
The information of most off-diagonal elements in the initial density matrix
cannot be extracted from the local observables after the system evolves into NESS.
In other words, they are forgotten in the evolution to NESS.
NESS only keeps memory of the off-diagonal elements with infinitesimal
energy difference.

We show that NESSH indicates ETH for the current operator in an arbitrary bipartite system.
And NESSH can be strictly proved in case of
a $2$-by-$2$ Hamiltonian by using the random matrix theory.
Furthermore, we provide the numerical
evidence of NESSH in two chaotic many-body models.
One composes of two weak-coupled reservoirs that are described by random matrices.
The other one is a lattice model of fermions.

According to our theory, in a generic system, i.e.
a system with realistic interactions between particles,
the NESS must be described by Eq.~(NESSH).
Just as the Gibbs ensemble is the general description of equilibrium states,
Eq.~(NESSH) is the general description of nonequilibrium steady states.
The Gibbs ensemble comes from the maximized entropy principle, but
Eq.~(NESSH) originates from quantum chaos theory.
NESSH is related to but distinguished from the eigenstate thermalization hypothesis.
The latter proposes an assumption about the observable operators
and uses it to explain why a system thermalizes.
Similarly, we propose an assumption about the density matrix and use it
to explain why a system evolves into NESS.
ETH is for a few-body observable operator, but the density matrix is not
a few-body operator.
More importantly, the structure $1/\left|E_\alpha-E_\beta\right|^2$ that we found
is absent in ETH.

The finding of a general structure in the nonequilibrium density matrix
is by itself awarding. Next we briefly discuss the possible applications of our theory.
First, the main difficulty in studying the NESS of nonintegrable models is that
the Schr\"{o}dinger equation is hard to solve. Especially, no numerical or analytical
approaches are reliable in the long time limit. Eq.~(NESSH) together
with~(\ref{eq:currsimplestexp}) provide an alternate
way. The steady current depends only upon the envelop function $\rho(\bar{E},\epsilon)$
in the diagonal limit $\epsilon\to 0$.
And calculating $\rho(\bar{E},0)$ is numerically economical,
since most information in the microscopic
wave function has no contribution to the steady current.
Second, our theory is useful in searching for general relations in the nonequilibrium states
since our formalism is independent of the model.
Finally, the form of density matrix given by Eq.~(NESSH) may help to
reveal the long-period decay of a finite system in the thermalization process.

\section*{Acknowledgements}

This work is supported by NSF of China under Grant No.~11304280.

\appendix
\section{Condition of a nonzero steady current}
\label{sec:app}

We start from the expression of current for an arbitrary initial state:
\begin{equation}
\begin{split}
 I(t)= \mathcal{C}_{\rho I} \int_{-\infty}^\infty d\bar{E} D(\bar{E}) \int_{-\infty}^\infty
 d\epsilon e^{-i\epsilon t} f_\rho (\bar E, \epsilon)
 f_I(\bar E, - \epsilon).
 \end{split}
\end{equation}
Here the time-dependent part is $\int d\epsilon e^{-i\epsilon t} f_\rho
 f_I$. A straightforward observation is that the asymptotic behavior of $I(t)$ at large $t$
depends on the asymptotic behavior of $f_\rho f_I$ at small $\epsilon$.
According to Riemann-Lebesgue lemma, if $f_\rho f_I$ is an integrable function,
$\int d\epsilon e^{-i\epsilon t} f_\rho f_I$ decays to zero as
$t\to \infty$. For the steady current being nonzero, $f_\rho f_I$ must diverge in the limit $\epsilon \to 0$. Let us
suppose that $f_\rho f_I$ diverges as $\Omega /\epsilon$ where $\Omega$
is a function of $\epsilon$ which converges to a finite value at $\epsilon = 0$.
The derivative of the current can be written as
\begin{equation}
 \frac{d I(t)}{dt}= -i \mathcal{C}_{\rho I} \int d\bar{E} D \int
 d\epsilon e^{-i\epsilon t} \left( \epsilon f_\rho
 f_I \right).
\end{equation}
Since $\epsilon f_\rho f_I=\Omega$ is an integrable function, $dI/dt$
must decay to zero as $t\to \infty$. This means that the current goes
to a stationary value in the long time limit. In fact, we have
\begin{equation}
\begin{split}
\lim_{t\to\infty} I(t) \sim & \lim_{t\to\infty} \int^\infty_{-\infty} d\epsilon e^{-i\epsilon t} \frac{\Omega(\epsilon)}{\epsilon} \\
 & =-i\pi \lim_{\epsilon\to 0}\Omega(\epsilon) ,
 \end{split}
\end{equation}
which is finite.

In general we should suppose that $f_\rho f_I$ diverges as $\Omega/\epsilon^\eta$.
By studying the Fourier transformation of $1/\epsilon^\eta$,
we find that $\displaystyle \lim_{t\to\infty}\int d\epsilon e^{-i\epsilon t}/\epsilon^\eta$ exists
and is nonzero if and only if $\eta =1$.


\begin{references}
\bibitem{laplace} The Laplace distribution with zero mean is defined as $P(x) \propto e^{-\left|x\right|/b}$.
\end{references}
\end{document}